**Accuracy of the Explicit Energy-Conserving Particle-in-Cell Method for Under-resolved Simulations of Capacitively Coupled Plasma Discharges**

A. T. Powis[1] & I. D. Kaganovich

*Princeton Plasma Physics Laboratory, Princeton NJ, USA*

**Abstract**

The traditional explicit electrostatic momentum-conserving Particle-in-cell algorithm requires strict resolution of the electron Debye length to deliver numerical accuracy. The *explicit* electrostatic *energy-conserving* Particle-in-Cell algorithm alleviates this constraint with minimal modification to the traditional algorithm, retaining its simplicity and ease of parallelization and acceleration on modern supercomputing architectures. In this article we apply the algorithm to model a one-dimensional radio-frequency capacitively coupled plasma discharge relevant to industrial applications. The energy-conserving approach closely matches the results from the momentum-conserving algorithm and retains accuracy even for cell sizes up to 8x the electron Debye length. For even larger cells the algorithm loses accuracy due to poor resolution of steep gradients in the radio-frequency sheath. This can be amended by introducing a non-uniform grid, which allows for accurate simulations with 9.4x fewer cells than the fully resolved case, an improvement that will be compounded in higher-dimensional simulations. We therefore consider the explicit energy-conserving algorithm as a promising approach to significantly reduce the computational cost of full-scale device simulations and a pathway to delivering kinetic simulation capabilities of use to industry.

**1. Introduction**

Delivering advanced semiconductor manufacturing capabilities for the coming decades will require an unprecedented control of plasma chemistry and kinetic behaviour at the wafer surface[1]. Developing equipment which can enable this control will necessitate increasing reliance on predictive modeling tools which can accurately capture this behaviour[2]. Such tools, perhaps in combination with artificial intelligence approaches, will also enable the generation of digital twins for plasma equipment and reduce the time for optimization of integrated circuit fabrication recipes[3,4].

Due to its intuitive nature, a long history of use within the community, and adaptability to high-performance computing, the Particle-in-Cell (PIC) method[5,6] has proven to be the most commonly used technique to conduct kinetic simulations of low-temperature plasmas, including low-pressure radio-frequency capacitively coupled plasma discharges (RF-CCPs) used in semiconductor etching, doping and cleaning[1,2,4,7,8,9,10,11,12,13,14,15,16,17]. In the electrostatic limit, the most common PIC algorithm solves for the Poisson equation, and therefore electric field, using the finite volume/difference method and updates the particles via the phase-space volume-preserving Boris algorithm[18,19]. When set on a uniform grid, this algorithm has the desirable property of reproducing momentum conservation up to machine precision[6] and we will therefore refer to this canonical algorithm as MC-PIC. Despite these properties, MC-PIC suffers from three numerical stability constraints:

1. Resolution of the electron Debye length: $\Delta x \lesssim \lambda_{De}$            (1a)

---

[1] Corresponding author: apowis@pppl.gov

2. Resolution of the electron plasma frequency: $\omega_{pe}\Delta t \lesssim 0.2$ (1b)
3. Satisfying the Courant condition for all particles: $v_{max} < \Delta x/\Delta t$ (1c)

Where $\Delta x$ is the cell size of the simulation, $\Delta t$ is the time step, $\lambda_{De}$ is the electron Debye length, $\omega_{pe}$ is the electron plasma frequency and $v_{max}$ is the velocity of the fastest simulation particle.

In this paper we will focus primarily on overcoming the restriction on cell size (Eq. 1a), although we will use the other stability constraints to draw conclusions regarding the convergence of our results. In simulations where the electron Debye length is underresolved, the instability, known as the finite-grid-instability, manifests as an artificial growth rate in the plasma wave dispersion relationship, leading to heating of the plasma[20]. This heating continues until condition 1a is satisfied, making it a particularly pernicious instability since the numerical scientist or engineer may mistake it for real physics within the simulation.

Perhaps unsurprisingly, these constraints make the computational cost of resolving the full six-dimensional phase space of a realistic plasma discharge enormous. Typical RF-CCP discharges may have dimensions on the order of thousands of $\lambda_{De}$ and equilibrium time scales spanning millions of plasma oscillations ($1/\omega_{pe}$). This has severely limited the use of computer aided software for designing industrial plasma devices which operate in the kinetic regime, necessitating costly and time-consuming experiments as an alternative. Advances in computing power have come a long way to alleviate these issues, however realistic fully resolved 2D simulations remain challenging and large 3D simulations are still impossible, even on the latest supercomputers, making them prohibitive for industrial computer aided design. Therefore, high-performance computing cannot be relied upon as the only means to address this challenge, research and development of new algorithms which can alleviate the stability constraints of Eq. 1 are also essential.

There have been several attempts to overcome the numerical constraints of the MC-PIC algorithm. These include semi-implicit methods such as the direct-implicit or implicit-moment methods which apply a single linearized correction and can be generalized to additional corrections[21,22,23,24,25,26,27,28,29,30,31,32,33,34,35,36,37]. These techniques can prove accurate when tuned correctly, however the optimal parameters to avoid numerical heating or cooling can change with time or space in the simulation, making them challenging to maintain[36,37]. More recently, work on fully implicit schemes has been driven by Chen & Chacon et al.[38,39,40,41,42,43], Lapenta et al.[44,45,46,47] and others[48], with Eremin et al.[16,49,50,51,52] adapting them for low-temperature applications. These approaches remain stable, and energy-conserving, for arbitrary cell and time step sizes, although some care must be taken with the particle update. While powerful, these techniques often require a fully non-linear solve with a well-tuned pre-conditioner, increasing code complexity and reducing parallel efficiency[53,54]. Additionally, many low-temperature plasma simulations are interested in the interaction between the plasma and the wall, which requires resolution of the sheath time scales ($1/\omega_{pe}$), rendering the ability to step over the plasma frequency less advantageous when modeling these discharges.

An alternative algorithm, attributable to a 1970 paper by Lewis[55], applies a variational approach to derive an explicit PIC method which is energy conserving in the zero-time-step limit. More recently these derivations have been put on a more theoretical footing through research into geometric PIC approaches[56,57,58]. In practice, the method tightly bounds energy growth when the time step satisfies stability condition 1b and for practical purposes can be considered energy conserving. Throughout this paper we will refer to this explicit electrostatic energy-conserving technique as EC-PIC. The advantage of energy conservation is that it prevents the finite-grid instability from occurring and therefore completely alleviates stability constraint 1a, allowing for $\Delta x \gg \lambda_{De}$. Additionally, EC-PIC is explicit, with only a slight modification to the MC-PIC method and is therefore easy to implement and highly amenable to parallelization and acceleration on modern heterogeneous supercomputing architectures. The EC-PIC algorithm was initially dismissed as infeasible due to the emergence of a non-physical cold-beam instability[59], however recent work from Barnes & Chacon has shown that the technique is accurate in thermal plasmas dominated by ambipolar forces[60]. More specifically, the Debye length can be safely underresolved when the velocity of a particle beam is less than the species thermal velocity. This makes the EC-PIC method a strong candidate for modeling many low-temperature plasma devices.

It is important to clarify that the EC-PIC approach alleviates a *numerical* constraint of the MC-PIC algorithm and will therefore not necessarily be accurate for $\Delta x > \lambda_{De}$ if resolution of Debye scale physics is required for *physical* accuracy. Therefore EC-PIC is ideal for modeling cases where $\lambda_{De}$ scale physics is unimportant throughout much of the plasma. An example of such a case is the physics occurring within a RF-CCP discharge, where the bulk of the plasma is quasi-neutral and relatively quiescent. For this reason, we choose to test EC-PIC on the industry relevant configuration of the RF-CCP discharge.

An additional caveat to consider is that the EC-PIC method does not conserve momentum to machine precision. Therefore, as with any numerical method, it is important to check the accuracy of EC-PIC in modeling the system of interest. In this paper we demonstrate that the approach can provide accuracy for RF-CCP simulations while adopting cell sizes much larger than the Debye length. Section 2 details the software written, and simulation configuration used to explore the accuracy of the method. Section 3 presents results comparing the MC-PIC and EC-PIC approaches in the fully resolved and underresolved cases, demonstrating an approach to improve EC-PIC accuracy for several underresolved cases and discussing the possible performance gains enabled by this approach. Finally, Section 4 offers some concluding thoughts and pathways towards realistic simulations of industrial devices using the EC-PIC algorithm.

## 2. Methodology

To avoid ambiguities of complex geometry and boundary conditions, and focus attention on the algorithm, the EC-PIC method is tested on a simple 1D RF-CCP model using the *mini-pic* code. *mini-pic* is a one-dimensional object-oriented Python package developed at the Princeton Plasma Physics Laboratory for rapid prototyping of new algorithms. All routines are written using the NumPy framework[61], which offers performance similar to that of compiler level languages. Although the code only runs on a single CPU all routines can be offloaded to a single GPU using the CuPy package[62]. *mini-pic* incorporates periodic or Dirichlet boundary conditions with a time varying RF potential and solves the Poisson equation via LU decomposition. Particles are updated with the leap-frog algorithm and Monte-Carlo collisions are implemented following the techniques of Vahedi & Surendra[63], with specific modifications suitable to match the benchmark of Turner et al.[7]. The code is benchmarked against all cases of Turner's 2013 RF-CCP benchmark, showing excellent agreement (results are presented in Appendix A).

### 2a. Explicit Energy-Conserving PIC Algorithm

The one-dimensional energy-conserving (EC)-PIC algorithm time loop on a uniform grid has the following steps[55]. Unless denoted otherwise the time step ($k$) is the same for all variables. The algorithm is quite similar to the momentum-conserving (MC)-PIC algorithm so we will clearly denote where the differences lie:

1. Interpolate the charge density to the grid using a first order shape function:

$$\rho_n = \frac{1}{\Delta x} \sum_s q_s \sum_i S_1\left(\frac{X_i - x_n}{\Delta x}\right) \tag{2}$$

$$S_1(x) = \begin{cases} 1 + x, & -1 \leq x < 0 \\ 1 - x, & 0 \leq x \leq 1 \end{cases} \tag{3}$$

Where $\rho_n$ is the charge density at grid node $n$ with location $x_n$. $X_i$ is the location of particle $i$ and $q_s$ is the charge of species $s$.

2. Solve the Poisson equation on the grid using the 3-point stencil for the Laplacian:

$$\frac{\phi_{n+1} - 2\phi_n + \phi_{n-1}}{\Delta x^2} = -\frac{\rho_n}{\varepsilon_0} \tag{4}$$

Where $\phi_n$ is the electric potential at grid node $n$. In this paper we set Dirichlet boundary conditions by assigning the potential at the end-point grid nodes $\phi_0$ and $\phi_{N+1}$.

3. (Different from MC-PIC). Rather than use a centered finite difference scheme, calculation of the electric field from the potential is given by a cell-centered finite volume approach:

$$E_n = -\frac{\phi_{n+1} - \phi_n}{\Delta x} \tag{5}$$

Where $E_n$ is the electric field at grid node $n$.

4. (Different from MC-PIC). Interpolation of the electric field back to the particles is performed via a shifted, non-symmetric, nearest grid point shape function:

$$E_i = \sum_n E_n S_0\left(\frac{X_i - x_n}{\Delta x}\right) \tag{6}$$

$$S_0(x) = \begin{cases} 1, & 0 \leq x < 1 \\ 0, & otherwise \end{cases} \tag{7}$$

Where $E_i$ is the electric field for particle $i$ and $S_0(x)$ is the offset nearest grid point shape function.

5. Update the particle phase space data $(X_i, V_i)$ via the leapfrog algorithm.

$$\frac{V_i^{k+1/2} - V_i^{k-1/2}}{\Delta t} = \frac{q_s}{m_s} E_i^k \tag{8}$$

$$\frac{X_i^{k+1} - X_i^k}{\Delta t} = V_i^{k+1/2} \tag{9}$$

6. Collide particles via the chosen Monte-Carlo Collision algorithms.
7. Return to Step 1.

Note that if we let $n \to n + 1/2$, for Eqs. (5)-(7), these can be interpreted as calculating the electric field on a staggered grid and calculation of the particle electric field via a centered, symmetric, nearest grid point shape function on that same staggered grid, however the authors prefer to avoid use of half-indices since they can more easily lead to confusion at implementation time. Higher order EC-PIC methods can be derived[56], however it is important that the shape function for interpolation of electric field to the particles remains one order lower than interpolation of charge density to the grid. Alternatively, one could update the electric field via Ampere's law and use the same shape function for current deposition and electric field interpolation[49], although in higher dimensions this approach would require a procedure to eliminate the rotational component of the updated electric field.

An empirical study of the heating rates for the MC-PIC and EC-PIC algorithms in a collisionless periodic system for different grid resolutions is presented in Appendix B, demonstrating the superior energy conservation properties of the EC-PIC method.

## 2b. Configuration of the Radio-Frequency Capacitively-Coupled Plasma discharge

To properly test the accuracy of EC-PIC for realistic low-temperature plasma physics the Helium RF-CCP simulations from Turner et al. are modified to that of a larger (or equivalently higher density) discharge. The simulation is initialized with a uniform plasma density and temperature, and evolved until steady state is achieved, after which the plasma properties are averaged over a given time period. We utilize identical Helium collision algorithms to those from the Turner benchmarks, including electron-neutral elastic collisions, two electron-neutral excitation collisions (at 19.82 eV and 20.61 eV), electron-neutral ionization (24.59 eV), ion-neutral elastic collisions and ion-neutral charge-exchange, with the same collision cross-section tables used by Turner et al.[7]. The physical and numerical parameters for the discharge are listed in Table 1.

**Table 1:** Parameters for the RF-CCP discharge simulations.

| Parameter Name | Parameter Symbol & Units | Parameter Value |
| --- | --- | --- |
| Electrode separation | $L\ (m)$ | 0.3 |
| Neutral density | $n_n\ (1/m^3)$ | $9.64 \times 10^{20}\ (30\ mTorr)$ |
| Neutral temperature | $T_n\ (K)$ | 300 |
| Frequency | $f\ (MHz)$ | 13.56 |
| Voltage amplitude | $V_0\ (V)$ | 1,000 |
| Simulation time | $T_{sim}\ (s)$ | $5,100/f$ |
| Total averaging time | $T_a\ (s)$ | $100/f$ |
| Averaging time interval | $T_f\ (s)$ | $1/10f$ |
| Initial plasma density | $n_0\ (1/m^3)$ | $7 \times 10^{15}$ |
| Initial electron temperature | $T_e\ (eV)$ | $2.585\ (30,000\ K)$ |
| Initial ion temperature | $T_i\ (K)$ | 300 |
| Reference number of cells | $N_{x,ref}$ | 2,048 |
| Reference cell size | $\Delta x_{ref}\ (m)$ | $1.465 \times 10^{-4}$ |
| Number of time steps | $N_t$ | 20,400,000 |
| Time step size | $\Delta t\ (s)$ | $1/4000f$ |
| Averaging time steps | $N_a$ | 400,000 |
| Averaging frequency time steps | $N_f$ | 400 |
| Initial particle-per-cell | $N_{PPC}$ | 512 |

All simulations were performed on a single NVIDIA V100 GPU with instructions from a single core of an IBM POWER9 CPU at Princeton Universities *Traverse* cluster[64] allowing for multiple simulations to be run concurrently.

## 3. Results & Discussion

### 3a. Comparison of Momentum & Energy Conserving PIC methods in the fully resolved case

The simulation is initialized and evolved for 5,000 RF periods or 20 million time steps until quasi-steady state is reached. Time averaged data of plasma moments, phase space data, and field quantities are then recorded over an additional 100 RF periods at intervals of $1/10^{th}$ of an RF period, or every 400 steps over a total 400,000 steps. At each output time all electron phase space data within a $1\ mm$ region at the center of the discharge are recorded for producing electron energy distribution functions (EEDFs). Additionally, all Helium ions which reach the RF electrode within this averaging time are recorded for generation of ion impact distribution functions. Temperature at each grid point is defined as the standard deviation of the species velocity distribution function as opposed to being related to the total kinetic energy. Figure 1 plots key plasma, field and energy distribution properties for both the fully resolved MC-PIC and EC-PIC simulations.

The electron and ion density exhibit the typical symmetric profile of an RF-CCP discharge with peak density at the center, decreasing monotonically towards the electrodes. Charge density reveals that the plasma is indeed quasi-neutral everywhere except within the sheaths near the electrodes. The potential shows a flat-topped profile, with peak plasma potential of 423 V, resulting in a near zero electric field in the plasma bulk tending to a large gradient in the sheath region. The x-component of electron temperature is raised above the initial simulation temperature to around 3.2 eV in the plasma bulk, peaking at the edge of the sheath where power is being deposited. The x-component of ion temperature is lowest in the bulk but increases in the sheath regions due to acceleration, although the idea of a temperature breaks down in this kinetic region. The electron energy distribution function (EEDF) at the center of the discharge show a typical bi-Maxwellian profile, with the higher energy tail important for driving ionization and generation of radicals in more complex chemistries[7,65]. The ion impact distribution functions exhibit a peaked profile corresponding to successive charge-exchange collisions during acceleration to the wall through the sheath potential[66]. The angular distribution is relatively narrow, with a minimum at $\theta = 0°$ caused by off-axis scattering during ion-neutral collisions in the sheath.

With consideration to the stability criterion of Eq. 1 we calculate that at quasi-stead state:

1. $\lambda_{De,min} \approx 1.11 \Delta x_{ref}$
2. $\omega_{pe,max} \Delta t \approx 0.085$
3. No particles exceed $v_{max} = \Delta x_{ref}/\Delta t$

Confirming that the simulation remains well resolved at steady-state.

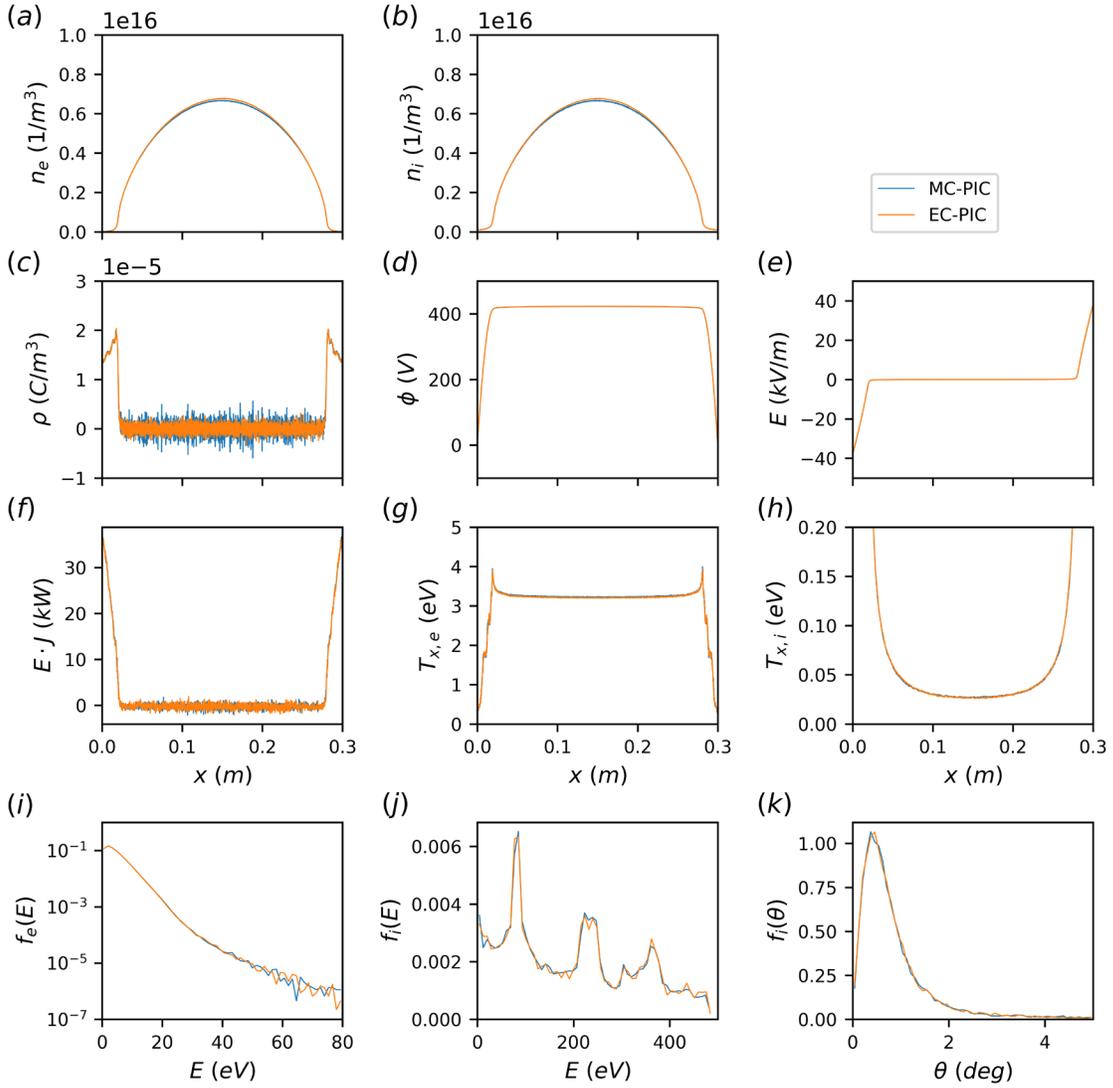

**Figure 1:** Time averaged plots for the fully resolved RF-CCP of (a) electron density, (b) ion density, (c) charge density, (d) potential, (e) electric field (f) power deposition, (g) x-component of electron temperature, (h) x-component of ion temperature, (i) EEDF at center of discharge, (j) ion impact energy distribution function and (k) ion impact angular distribution function. The momentum-conserving (MC)-PIC and energy-conserving (EC)-PIC methods show excellent agreement.

Figure 1 also demonstrates excellent agreement between the MC-PIC and EC-PIC algorithms. With the 1-norm difference in plasma density and electron temperature across the entire domain calculated as 1.1% and 0.76% respectively. Most importantly for industrial purposes the central electron EEDFs and ion impact energy and angular distribution functions also show close agreement (note that a 1-norm comparison here is somewhat superfluous since it will depend on the histogram function settings). Assuming that the well-resolved MC-PIC implementation is accurate due to it satisfying the PIC constraints, we can conclude that both algorithms accurately reproduce the behaviour of the model RF-CCP discharge.

## 3b. Stability and accuracy for under resolved simulations

To determine the stability and accuracy of the two algorithms for under resolved cases we progressively increase the cell size by powers of 2, up to a maximum cell size of $\Delta x = 32 \Delta x_{ref}$, corresponding to $N_x = 64$ simulation cells. For all simulations the total number of initial particles is kept the same. Figure 2 and Fig. 3 plot the same parameters as Fig. 1 for the MC-PIC and EC-PIC algorithms respectively, with progressively increasing cells size.

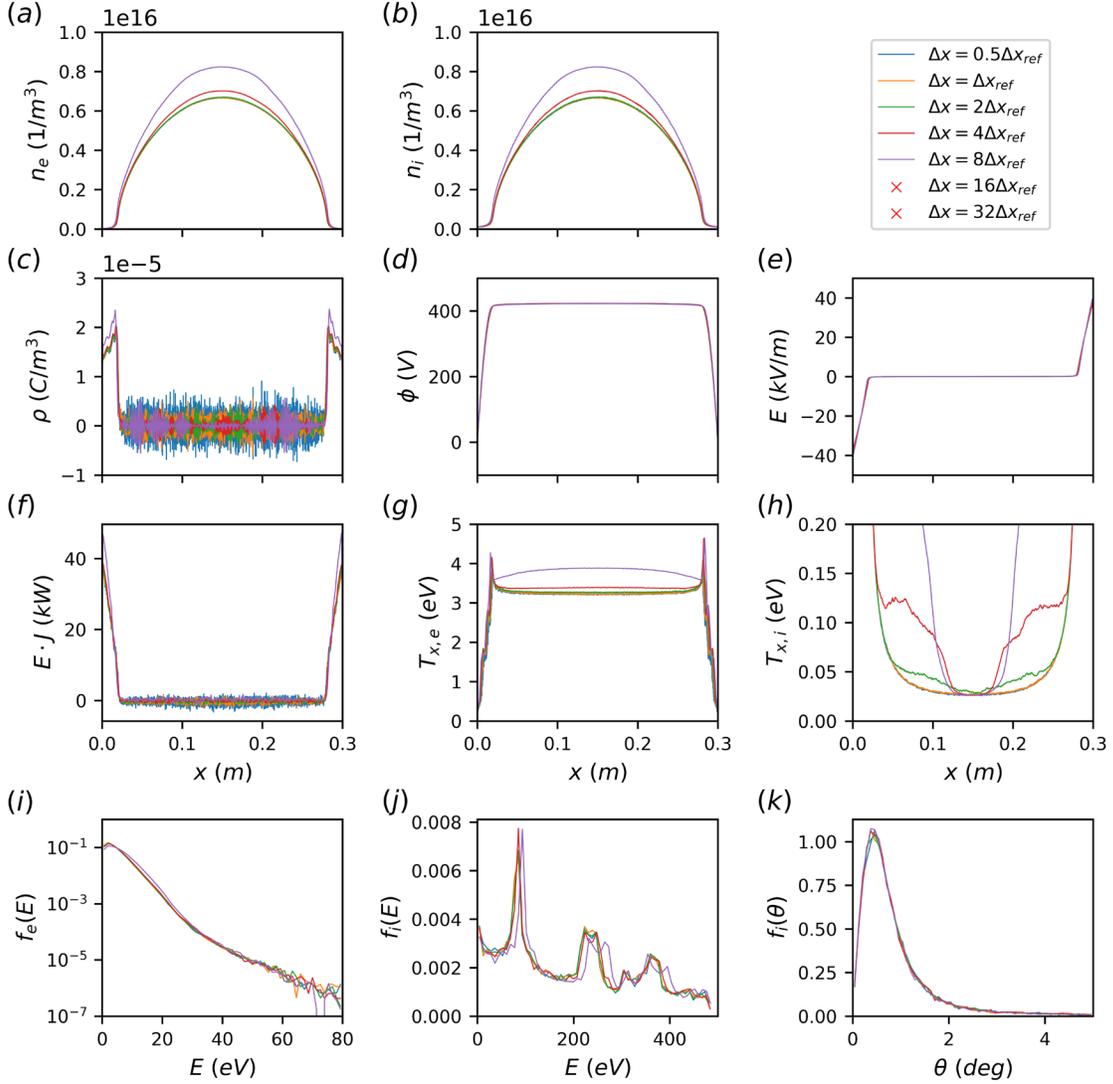

**Figure 2:** Time averaged plots for the RF-CCP with different cell sizes using the momentum-conserving (MC)-PIC algorithm. Plots of (a) electron density, (b) ion density, (c) charge density, (d) potential, (e) electric field (f) power deposition, (g) x-component of electron temperature, (h) x-component of ion temperature, (i) EEDF at center of discharge, (j) ion impact energy distribution function and (k) ion impact angular distribution function. MC-PIC is unstable for $\Delta x > 8\Delta x_{ref}$ and inaccurate for $\Delta x > 2\Delta x_{ref}$.

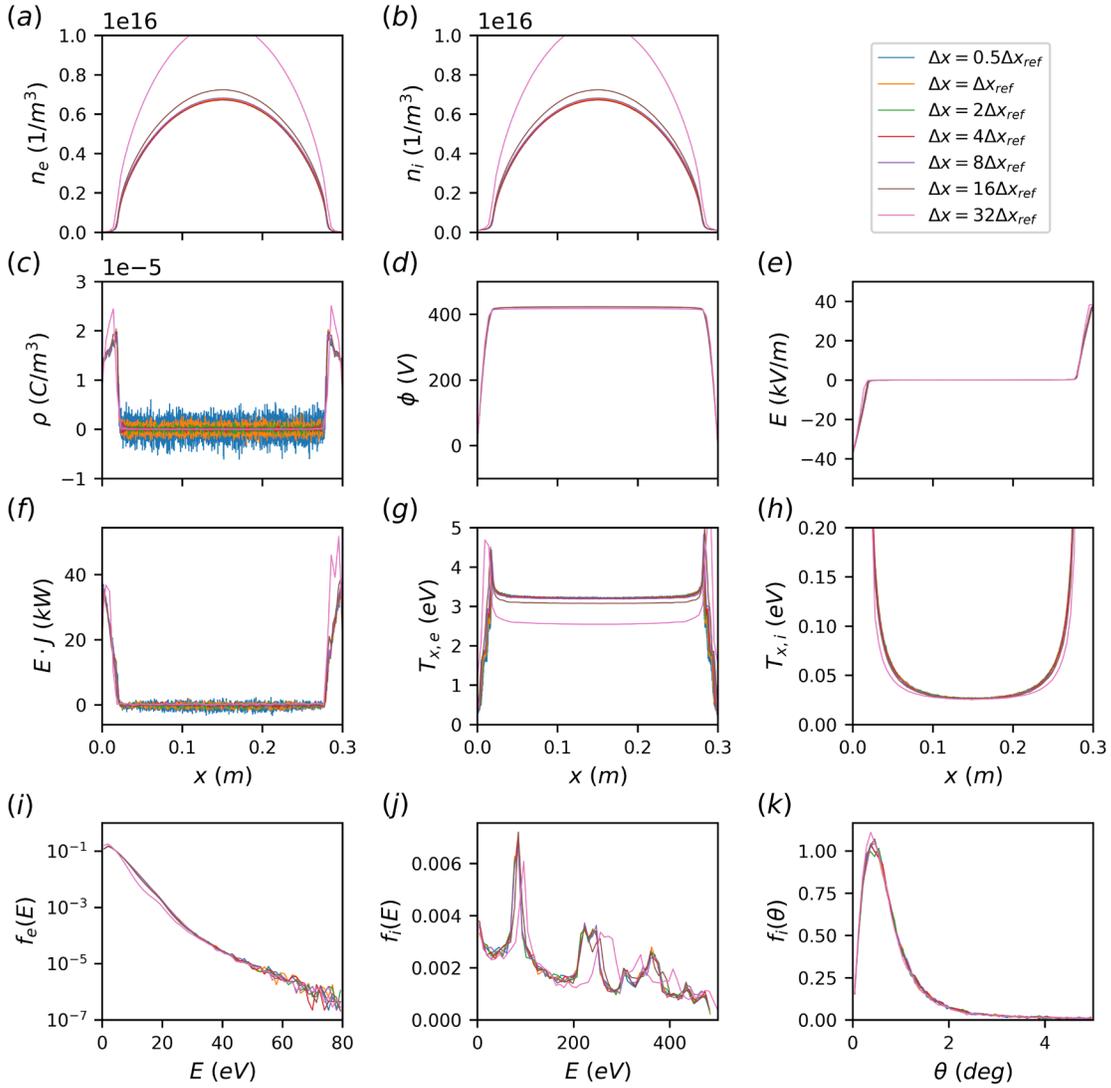

**Figure 3:** Time averaged plots for the RF-CCP with different cell sizes for the energy-conserving (EC)-PIC algorithm. Plots of (a) electron density, (b) ion density, (c) charge density, (d) potential, (e) electric field (f) power deposition, (g) x-component of electron temperature, (h) x-component of ion temperature, (i) EEDF at center of discharge, (j) ion impact energy distribution function and (k) ion impact angular distribution function. EC-PIC is stable for all cases, but inaccurate for $\Delta x > 8\Delta x_{ref}$ or $\Delta x \gtrsim 1\,mm$.

For the MC-PIC algorithm the simulations for $\Delta x/\Delta x_{ref} = \{16, 32\}$ become unstable and crash after 139 and 56 RF periods respectively (denoted by the red crosses in the Fig. 2 legend). Investigating the data reveals that the crash occurs when the 40 GB of GPU RAM becomes saturated. This indicates a 100 fold increase in memory utilization from the initial conditions of the simulation which, based on the code design, can only be attributed to creation of new particle phase space data. This particle creation is presumed to be caused by artificial heating from the finite-grid-instability due to under resolution of the Debye length, leading to an increase in ionization and a corresponding increase in density. This is then exacerbated when the density yields a plasma frequency too large for resolution by $\Delta t$ leading to an additional numerical instability and therefore further heating and ionization. Eventually this process leads to a runaway in particle number, memory exhaustion and finally simulation failure.

Regarding the stable results, we observe that for increasing $\Delta x$ the peak plasma density and electron temperature increase. The ion temperature increases towards the sheath region for the lowest resolution simulation ($\Delta x = 8\Delta x_{ref}$), this likely explains the shift to higher energies of the ion impact energy distribution function for the same case. In summary, the MC-PIC algorithm appears to become inaccurate for $\Delta x > 2\Delta x_{ref}$.

Contrasting this, the EC-PIC simulations remain stable for all cases, presumably since the finite-grid instability is suppressed by this algorithm. Although stable, the density does increase significantly in the coarsest simulation, with a surprising drop in electron temperature in the plasma bulk but increase in the sheath region, driving additional ionization and therefore plasma density. The ion temperatures however remain extremely accurate for all cases and the EEDF and ion energy distribution functions are accurate for all but the coarsest case. In summary, the EC-PIC algorithm is stable for all cases, but inaccurate for $\Delta x > 8\Delta x_{ref}$ (approximately 1 mm), offering a significant improvement over the MC-PIC algorithm.

It should also be pointed out that decreasing the number of cells reduces the noise for all results, especially noticeable in the charge density (Fig. 2c and 3c). This is due to the fact that the total number of particles at initialization within each simulation remain the same, therefore the number of particles-per-cell increase with increasing cell size leading to a reduction in noise.

To investigate why the EC-PIC algorithm loses accuracy Fig. 4 plots the charge density and electric field within the left-hand sheath region ($x < 3cm$) at various phases of the RF bias for the different resolution simulations. It is clear that the gradient of charge density and electric field become quite steep during certain phases of the RF period, particularly at $4/(5f)$, and the simulations with $\Delta x > 8\Delta x_{ref}$ struggle to resolve this.

The charge density data is smoothed with a Savitzky-Golay filter[67] using a 2nd order polynomial and 21 neighbouring grid nodes (Fig. 5a). The smoothed data ($\rho_s$) is used to calculate the gradient length scale of the electric field as,

$$L_E = \frac{E}{dE/dx} = \varepsilon_0 \frac{E}{\rho_s} \tag{10}$$

Figure 5b plots $L_E$ normalized by $\Delta x_{ref}$ at the $4/(5f)$ phase for the fully resolved case, along with dashed lines corresponding to the cell sizes for each simulation. The gradient length scale is just under resolved for the case $\Delta x = 8\Delta x_{ref}$, although this case does accurately reproduce the RF-CCP behaviour, perhaps because the gradient length scale is appropriately resolved during most of the RF period. However, $L_E$ is clearly very under resolved for $\Delta x > 8\Delta x_{ref}$ which may explain the poor accuracy with these larger cell sizes. These results suggest that adopting a non-uniform grid to better resolve the dynamics within the sheath, while still allowing for larger cell sizes in the center, may help to remedy the accuracy issues of the EC-PIC method. Note that beyond the sheath region ($x > 2.2\ cm$) the data becomes noisy due to $\rho_s \to 0$, therefore the gradient length scale becomes meaningless in this region.

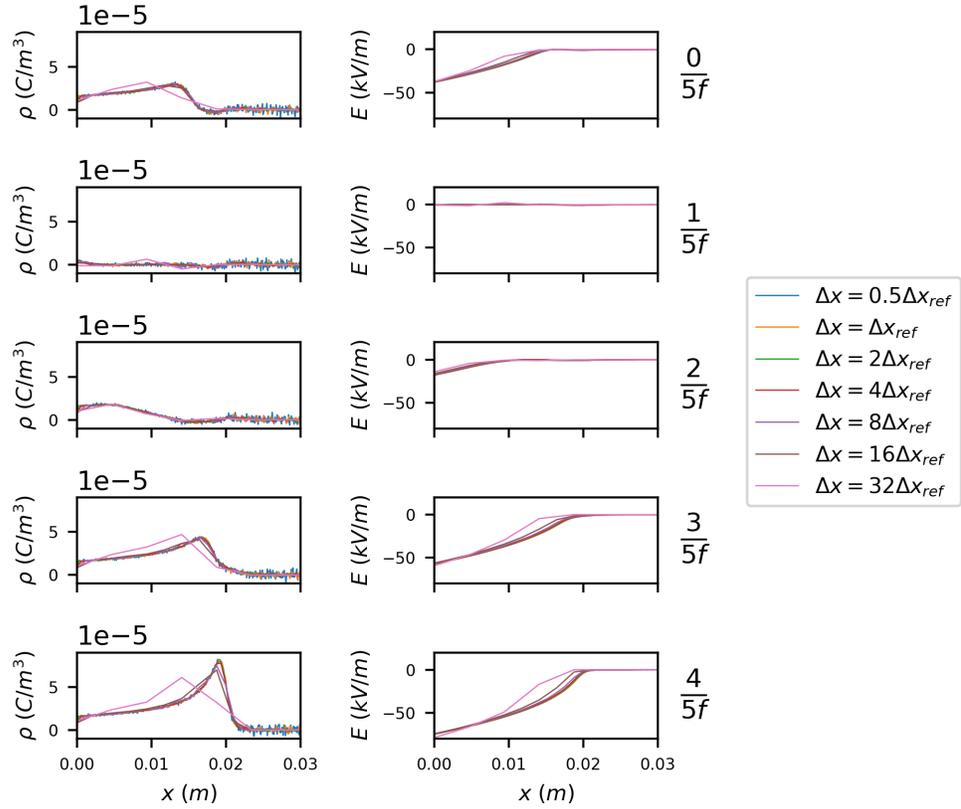

**Figure 4:** Time averaged plots at $1/5^{th}$ phases of the RF cycle within the left sheath region of the RF-CCP discharge with the energy-conserving (EC)-PIC algorithm for different cell sizes. Plots of (a) charge density, (b) electric field. The steepest gradients are poorly resolved for $\Delta x > 8\Delta x_{ref}$.

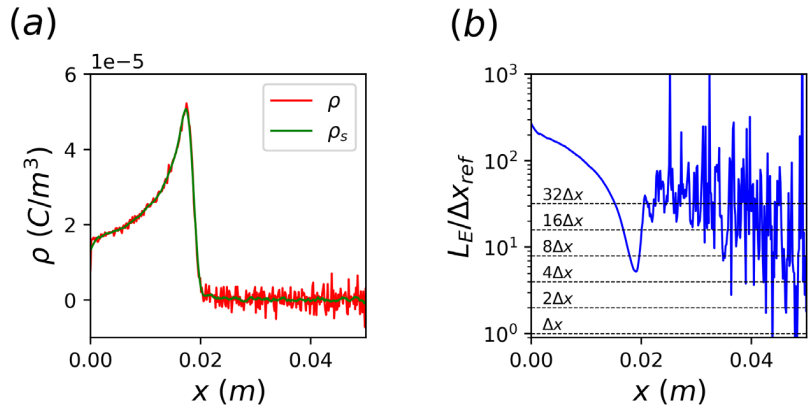

**Figure 5:** Time averaged plots at the $4/(5f)$ phase of the RF cycle within the sheath region of the RF-CCP discharge with the energy-conserving (EC)-PIC algorithm for the full resolution case. Plots of (a) charge density with data smoothed by a Savitzky-Golay filter and (b) electric field gradient length scale. The coarsest grid resolution ($\Delta x > 8\Delta x_{ref}$) cannot accurately resolve the smallest gradient length scales.

## 3c. Improving accuracy with non-uniform grids

Non-uniform grids introduce additional computational and accuracy challenges for PIC codes. In one-dimension the ordering of cells is straightforward, however higher dimensions raise questions on how to best order the cells for optimal memory access and communication patterns. Although not-essential for non-uniformity, an un-structured grid may also prevent the use of algorithms such as the *geometric* multigrid method, which often perform better than the more generalized *algebraic* multigrid method for solving the Poisson equation. Finally, non-uniform grids also introduce accuracy errors due to the introduction of a particle self-force, which breaks momentum conservation (even for MC-PIC)[68,69]. It therefore remains important to proceed with accurate numerical experiments to characterize the accuracy of RF-CCP simulations under various non-uniform grid conditions.

With this in mind, *mini-pic* was updated to incorporate non-uniform grids, with the following grid design logic:

1. Within a buffer length $L_{buff}$ from each electrode the cell size is set to $\Delta x_{min}$ such that sheath gradients are suitable resolved.
2. Outside of this region the cells can increase up to $\Delta x_{max}$.
3. Cells are increased (or decreased) in size by doubling (or halving) the cells size.
4. No cell can have neighbor cells which are larger (or smaller) by more than a factor of 2.

An example of how this refinement strategy might look is shown in Fig. 6 for $N_x = 128$, $L_{buff} = 10\Delta x_{ref}$, $\Delta x_{min} = \Delta x_{ref}$ and up to $\Delta x_{max} = 8\Delta x_{ref}$.

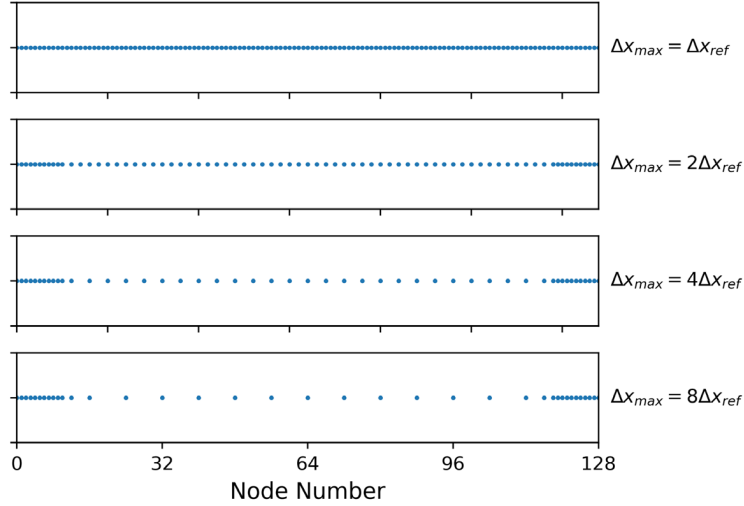

**Figure 6:** Example of the grid refinement strategy used for RF-CCP discharge simulations. $N_x = 128$, $L_{buff} = 10\Delta x_{ref}$, $\Delta x_{min} = \Delta x_{ref}$ and $\Delta x = [1,2,4,8]\Delta x_{ref}$.

From Fig. 4 we observe that the maximum sheath length during the RF phase is approximately 2.2cm, and therefore set $L_{buff}$ in our RF-CCP simulations to be just over two times the sheath length or 5 cm. The total number of cells for each simulation are shown in Table 2. Figure 7 plots the results of the EC-PIC simulations with the refined grid and increasing maximum cell sizes. For all cases the same number of total simulation particles are used.

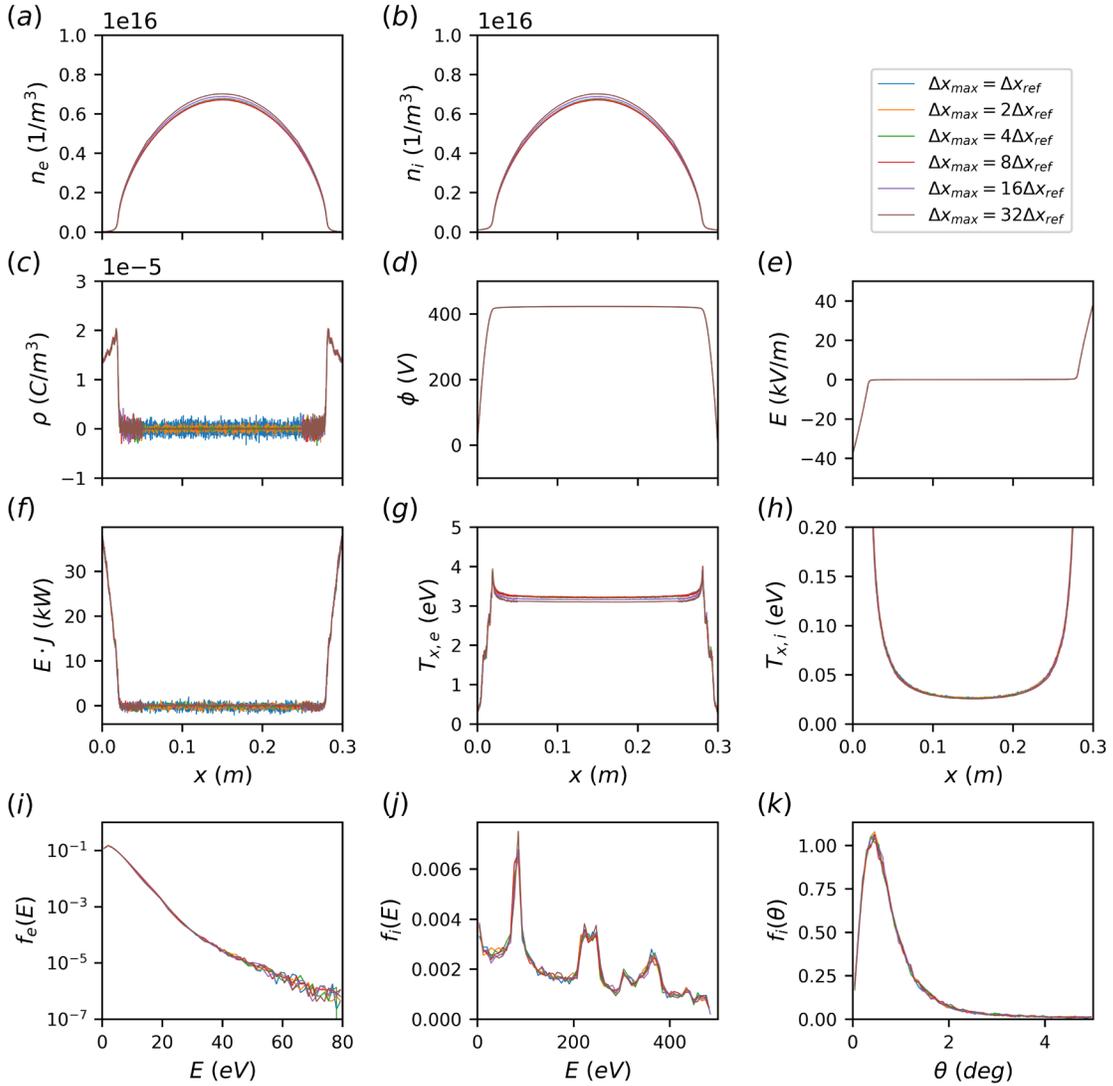

**Figure 7:** Time averaged plots for the RF-CCP with non-uniform grids with different maximum cell sizes for the energy-conserving (EC)-PIC algorithm. Plots of (a) electron density, (b) ion density, (c) charge density, (d) potential, (e) electric field (f) power deposition, (g) x-component of electron temperature, (h) x-component of ion temperature, (i) EEDF at center of discharge, (j) ion impact energy distribution function and (k) ion impact angular distribution function. All cases accurately reproduce the results from the high resolution fully resolved case.

We clearly observe a significant increase in accuracy for the larger cell size cases when compared to the uniform grid cases. Although the minimum cell size has been set to the reference cell size in Fig. 7, the accuracy of EC-PIC in the results of Fig. 3 indicates that we may be able to safely adopt larger minimum cell sizes as well. This is due to the fact that the sheath can remain accurately resolved at these larger cell sizes as shown by Fig. 6. Figure 8 shows additional simulations with different minimum and maximum cell sizes.

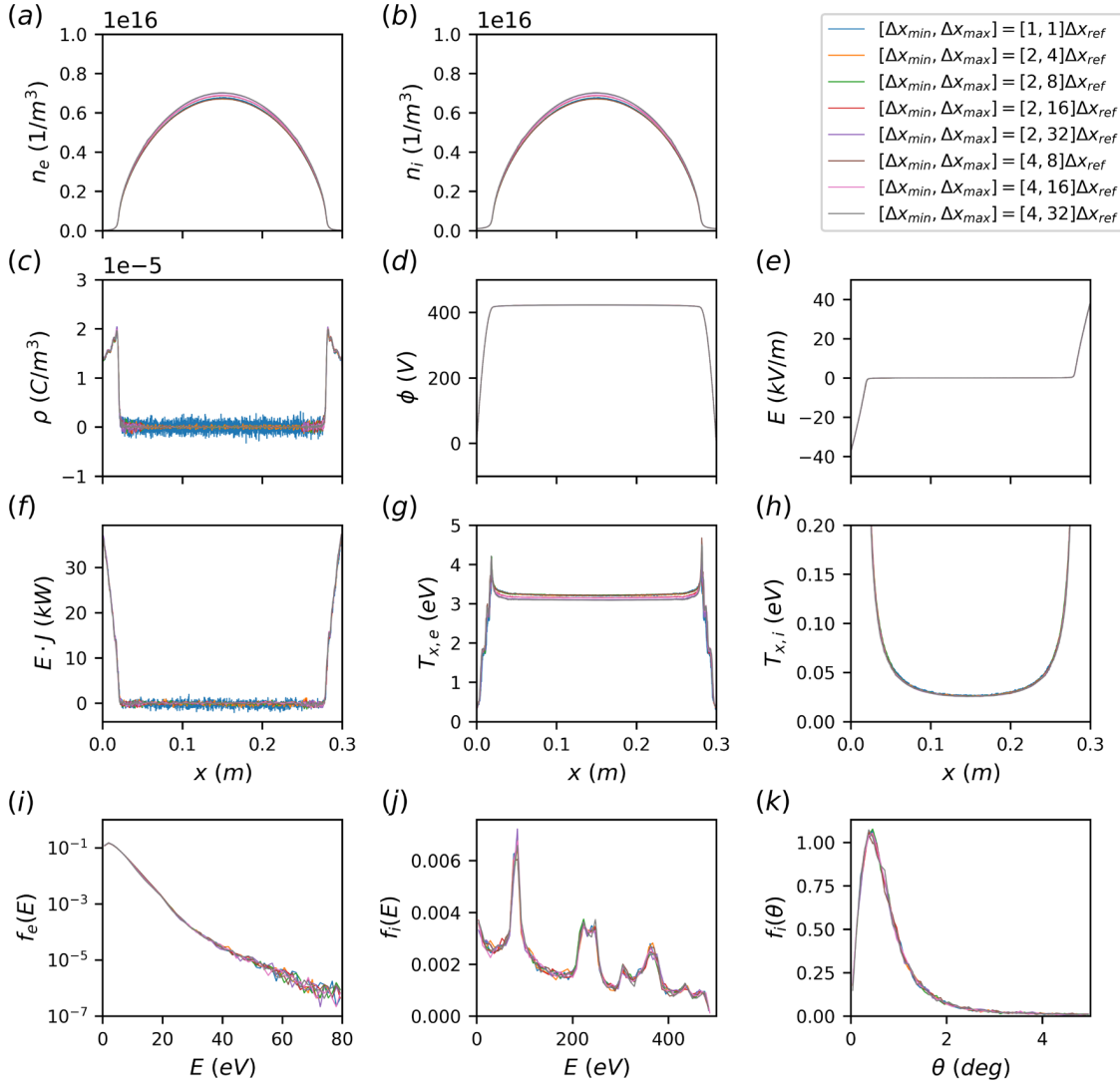

**Figure 8:** Time averaged plots for the RF-CCP with non-uniform grids with different minimum and maximum cell sizes for the energy-conserving (EC)-PIC algorithm. Plots of (a) electron density, (b) ion density, (c) charge density, (d) potential, (e) electric field (f) power deposition, (g) electron temperature, (h) ion temperature, (i) electron energy distribution function at center of discharge, (j) ion impact energy distribution function and (k) ion impact angular distribution function. All cases accurately reproduce the results from the high resolution fully resolved case.

Table 2 summarizes the 1-norm discrepancy between the different minimum and maximum cell size cases. For mismatched grid sizes we use a 1D interpolation function to map the coarse data onto the unrefined reference grid with $\Delta x = \Delta x_{ref}$.

**Table 2:** Total number of cells and 1-norm of discrepancy between plasma density and electron temperature for different grid refinement minimum and maximum cell sizes. The reference case for all comparisons is with $\Delta x_{min} = \Delta x_{max} = \Delta x_{ref}$ (first row).

| Minimum cell size | Maximum cell size | Total number of cells | Ion density error (1-norm) | Electron temperature error (1-norm) |
|---|---|---|---|---|
| $\Delta x_{ref}$ | $\Delta x_{ref}$ | 2,048 | 0% | 0% |
| $\Delta x_{ref}$ | $2\Delta x_{ref}$ | 1,366 | 0.34% | 0.45% |
| $\Delta x_{ref}$ | $4\Delta x_{ref}$ | 1,026 | 0.62% | 0.49% |
| $\Delta x_{ref}$ | $8\Delta x_{ref}$ | 858 | 0.64% | 0.71% |
| $\Delta x_{ref}$ | $16\Delta x_{ref}$ | 776 | 2.1% | 2.0% |
| $\Delta x_{ref}$ | $32\Delta x_{ref}$ | 736 | 4.2% | 3.7% |
| $2\Delta x_{ref}$ | $2\Delta x_{ref}$ | 1,024 | 0.46% | 0.90% |
| $2\Delta x_{ref}$ | $4\Delta x_{ref}$ | 684 | 0.58% | 0.87% |
| $2\Delta x_{ref}$ | $8\Delta x_{ref}$ | 516 | 0.58% | 1.1% |
| $2\Delta x_{ref}$ | $16\Delta x_{ref}$ | 434 | 2.1% | 2.5% |
| $2\Delta x_{ref}$ | $32\Delta x_{ref}$ | 394 | 3.9% | 4.1% |
| $4\Delta x_{ref}$ | $4\Delta x_{ref}$ | 512 | 0.52% | 1.4% |
| $4\Delta x_{ref}$ | $8\Delta x_{ref}$ | 342 | 0.65% | 1.7% |
| $4\Delta x_{ref}$ | $16\Delta x_{ref}$ | 258 | 2.5% | 3.4% |
| $4\Delta x_{ref}$ | $32\Delta x_{ref}$ | 218 | 4.3% | 5.1% |

Using an accuracy benchmark of 5% we can conclude that for almost all cases the EC-PIC method accurately reproduces the RF-CCP behaviour. The only exception is the case where $\Delta x_{min} = 4\Delta x_{ref}$ and $\Delta x_{max} = 32\Delta x_{ref}$ (the most under-resolved case) where the 1-norm of electron temperature slightly exceeds this benchmark at 5.1%. Perhaps most importantly for applications to semi-conductor etching the EEDF at the center of the discharge, ion impact distribution energy and angle are clearly in excellent agreement for all cases. This shows that in our configuration the RF-CCP discharge can be accurately modeled using the EC-PIC approach on non-uniform grids, providing an up to 9.4x reduction in the number of cells. Appendix C shows similar results for the MC-PIC algorithm demonstrating that accuracy is not improved when a similar non-uniform grid approach is adopted. The next section will discuss how this can lead to improved performance for one and higher dimensional CCP simulations.

**3d. Estimated performance improvements**

With consideration to the minimum cell size $\Delta x_{min}$, maximum cell size $\Delta x_{max}$ and the buffer length $L_{buff}$ the number of cells on the refined grid can be estimated as,

$$N_x = \frac{L - 2L_{buff}}{\Delta x_{max}} + \frac{2L_{buff}}{\Delta x_{min}} \quad (11)$$

Where we have ignored the small transition region in cell size from $\Delta x_{min}$ to $\Delta x_{max}$. Normalizing by the number of cells in an unrefined grid we can compute the reduction in number of cells ($CR$),

$$CR = 2\left(1 - \frac{1}{R}\right)\frac{L_{buff}}{L} + \frac{1}{R} \qquad (12)$$

Where $R = \Delta x_{max}/\Delta x_{min}$ is the grid refinement ratio.

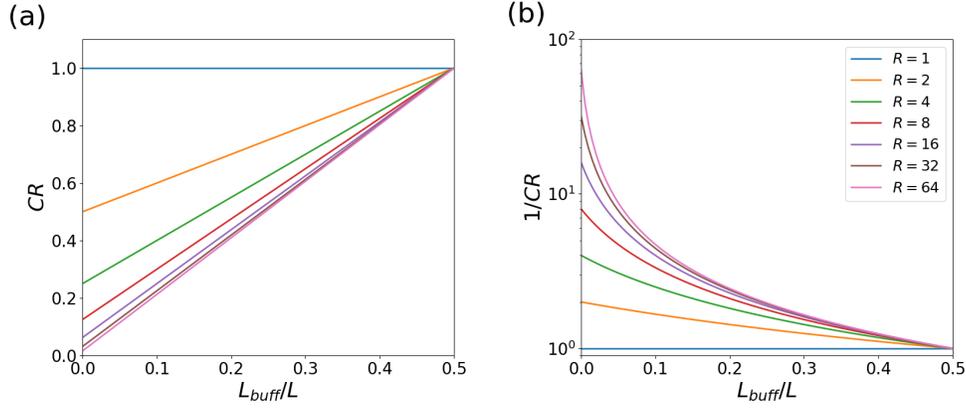

**Figure 9:** (a) Reduction in number of cells $CR$ and (b) $1/CR$ against ratio of buffer length to simulation size for various grid refinement ratios. With appropriate particle balancing algorithms $1/CR$ may be considered an estimate for the simulation speedup in one-dimension.

Figure 9a shows $CR$ against $L_{buff}/L$ for various refinement ratios demonstrating that, unsurprisingly, the reduction in number of cells is linearly proportional to $L_{buff}/L$ and decreases asymptotically towards zero for $L_{buff}/L = 0$ as $R \to \infty$. As stated in Section 3b and 3c all simulation are run with the same total number of particles, however let us consider a hypothetical code which has both non-uniform grids and an algorithm to periodically rebalance the number of particles to maintain a relatively constant number of particles-per-cell. For such an algorithm the field solve time and particle update time are directly proportional to the number of cells and we can therefore consider $1/CR$ as an estimate of the simulation speedup in one-dimension. With these assumptions in mind Fig. 9b reveals some interesting trends, firstly, increasing the refinement ratio leads to the expected improvements in performance. Secondly the benefits of grid refinement diminish rapidly as $L_{buff}$ approaches $L$, indicating that EC-PIC with non-uniform grids may only offer improvements if a large portion of the simulation length can be safely under-resolved. However, when the fully-resolved region is small compared to the plasma bulk the non-uniform grid approach with particle rebalancing can offer a significant speedup. Furthermore, if we assume that the problem length scales are similar in each direction this speedup will scale as $(1/CR)^d$, where $d$ is the dimension of the problem. For the coarsest grid case from Fig. 8 the speedup could be 88x in 2D and 830x in 3D. The authors are currently investigating the accuracy of various particle rebalancing algorithms and will report the findings of this work in a future publication.

Even without an appropriate particle-rebalancing algorithm, when paired with an appropriate field solver, the reduced number of cells can offer improved field solver performance. This can become particularly important for GPU accelerated PIC codes, where the improved acceleration of the particle routines can expose reduced performance of the field solver.

## 4. Conclusion

A comparison is made between the explicit momentum conserving and explicit energy conserving PIC algorithms through modeling of an RF capacitively coupled plasma discharge. It is observed that the EC-PIC method can accurately reproduce the behaviour of the discharge with respect to the MC-PIC simulations. When under resolving the simulation with larger cell size the EC-PIC method is observed to be more stable and accurate that the MC-PIC case. Loss in accuracy of the EC-PIC method for highly underresolved simulations is due to poor resolution of the plasma sheath region which can be remedied by adopting a non-uniform grid. The non-uniform EC-PIC method shows both excellent accuracy and stability, allowing for a reduction in number of cells by up to a factor of 9.4. This will lead to reduced computational cost for PIC grid-based routines and if combined with appropriate particle rebalancing schemes may significantly reduce overall computational cost, with even greater benefit in two or three-dimensional simulations. Future work will investigate how best to implement such particle rebalancing schemes as well as demonstrate the approach in higher-dimensions. This work presents one step towards realizing PIC as a useful tool for engineering prototyping of low-temperature plasma devices relevant to a wide range of industries.

**Acknowledgements**

This research was funded by the Department of Energy's Laboratory Directed Research and Development (LDRD) program.

**Data Availability Statement**

The data that support the findings of this study are available from the corresponding author upon reasonable request.

## Appendix A: Benchmarking of *mini-pic*

*mini-pic* was benchmarked against Turner et al.'s 2013 RF-CCP simulations which provide plots of plasma density for comparison. Results from *mini-pic* and the benchmark are shown in Figure A1 demonstrating excellent agreement for all cases. Based on this data we consider *mini-pic* an accurate PIC-MCC code for the purposes of this paper.

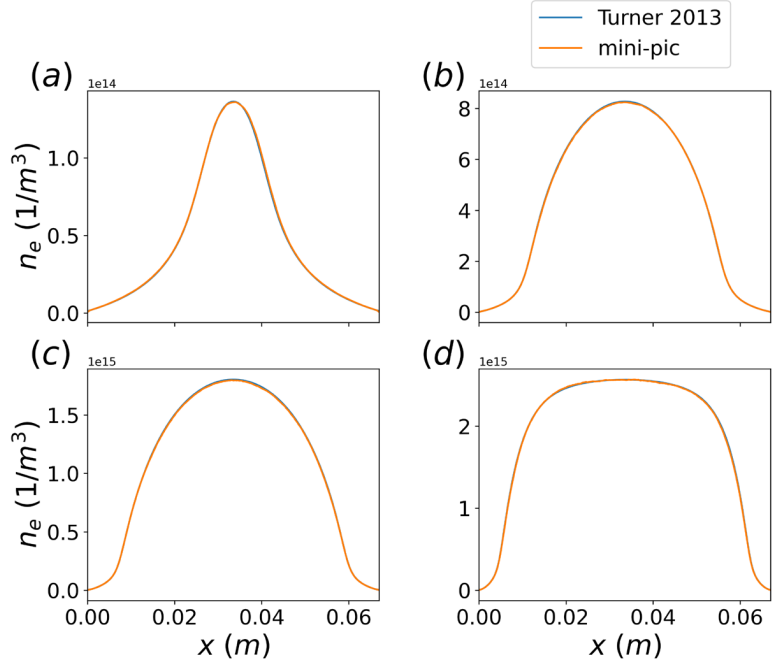

**Figure A1:** Comparison of time averaged electron density between the *mini-pic* code and all cases of Turner's 2013 benchmark of RF-CCPs[7]. (a) case 1, (b), case 2, (c) case 3, (d) case 4. Where the blue line is not visible it is completely overlapped by the orange line.

## Appendix B: Empirical heating rates of the explicit momentum and energy conserving PIC algorithms

To demonstrate the excellent energy conserving properties of EC-PIC for underresolved simulations we model $10{,}000/2\pi$ plasma periods on a grid with 2,048 cells using time step $\Delta t = 0.1/\omega_{pe}$. The boundary conditions are periodic and no collisions are included, so the initial number of particles and total energy should remain constant for an accurate simulation. The length of the simulation is increased such that the cell size scales according to the reference electron Debye length given by the initial electron density and temperature.

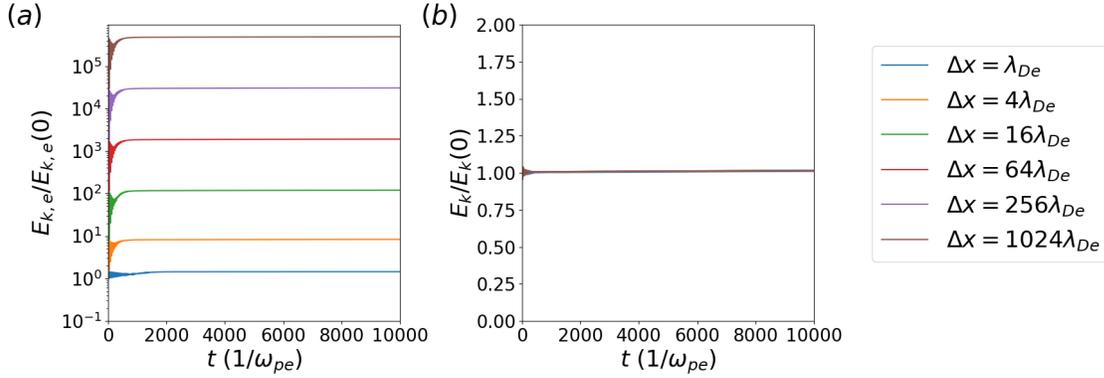

**Figure B1:** Change in relative electron kinetic energy (or temperature) for (a) momentum-conserving (MC)-PIC and (b) energy-conserving (EC)-PIC with different grid resolutions. The increase in kinetic energy for MC-PIC is due to "grid-heating" as a result of the finite-grid-instability.

It is clear that for the MC-PIC algorithm the finite-grid-instability significantly heats the plasma until stability criterion 1a is satisfied, which leads to large numerical errors in under-resolved simulations. For the EC-PIC case however the error is tiny, even for highly under-resolved simulations. Despite these results the user should take care to check how under-resolution of Debye scale physics or lack of momentum-conservation effects their specific configuration before relying completely on EC-PIC.

**Appendix C: Momentum Conserving PIC with non-uniform grids**

Note that to call the momentum-conserving (MC)-PIC algorithm momentum conserving when using a non-uniform grid is a misnomer, since the non-uniform nature of the cells introduces particle self-forces that compromises momentum conservation. However, we continue to refer to it as such in this section to improve continuity.

To verify that the improvements in accuracy for the RF-CCP simulations with EC-PIC and a non-uniform grid are a result of *both* changes, it is important to see if the accuracy of MC-PIC is significantly improved on a non-uniform grid. Figure C1 shows similar plots of the RF-CCP discharge as those within the main body of the text. Although we do see an increase in accuracy with this approach, likely due to part of the simulation domain being appropriately resolved to satisfy stability criteria 1a, we still see that the method is inaccurate for $\Delta x > 4\Delta x_{ref}$ and fails for $\Delta x > 8\Delta x_{ref}$. This indicates that is not the introduction of a non-uniform grid alone which produces more accurate simulations.

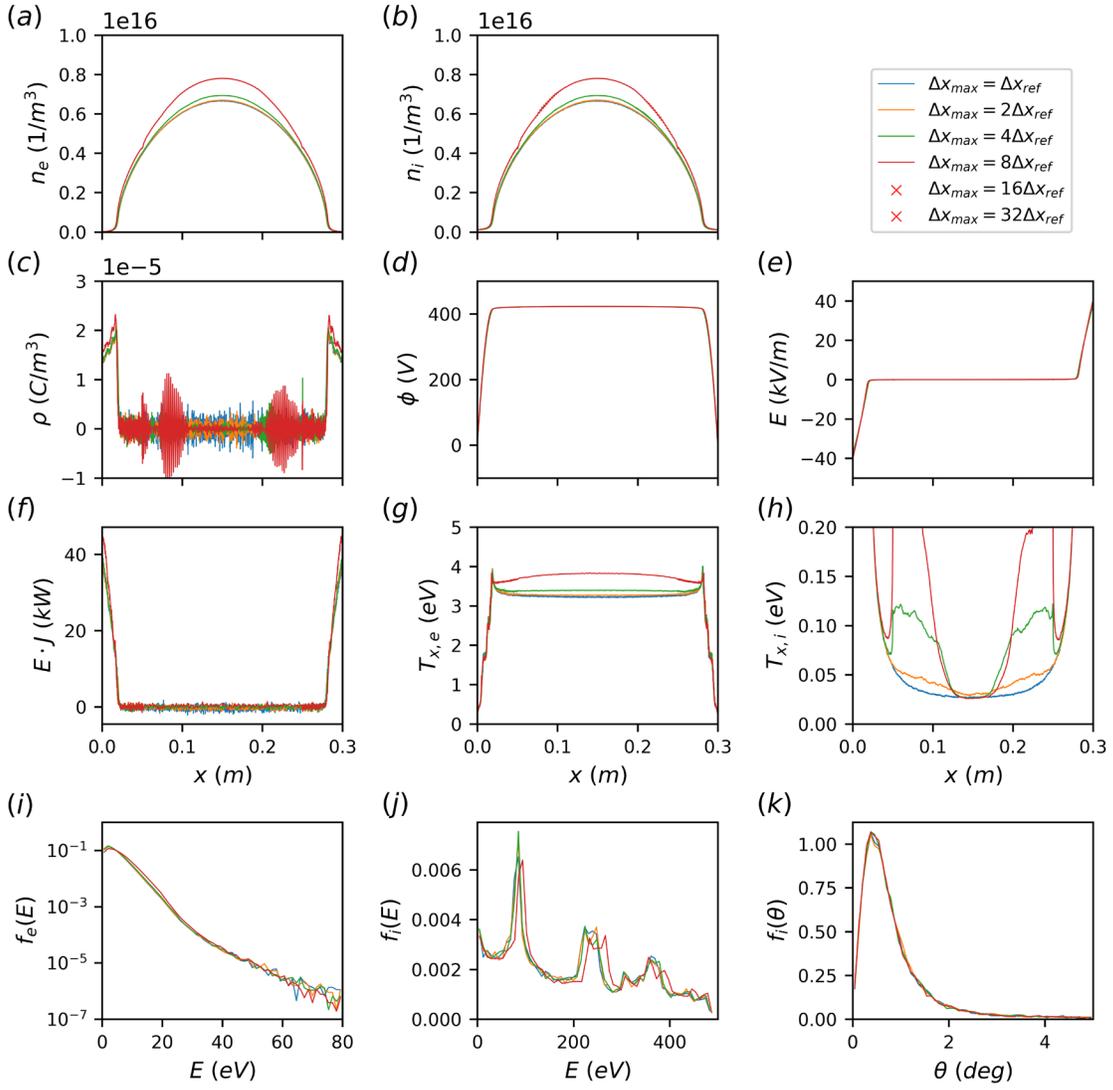

**Figure C1:** Time averaged plots for the RF-CCP with different cell sizes using the momentum-conserving (MC)-PIC algorithm. Plots of (a) electron density, (b) ion density, (c) charge density, (d) potential, (e) electric field (f) power deposition, (g) x-component of electron temperature, (h) x-component of ion temperature, (i) EEDF at center of discharge, (j) ion impact energy distribution function and (k) ion impact angular distribution function. MC-PIC is unstable for $\Delta x > 8\Delta x_{ref}$ and inaccurate for $\Delta x > 4\Delta x_{ref}$.